\documentstyle[11pt,newpasp,twoside,epsf]{article}
\def\gapprox{{_>\atop{^\sim}}}

\def\cmmt{\rm {cm^{-2}}}

\def\s-1{\rm {s^{-1}}}
\def\twco{$^{12}$CO}
\def\thco{$^{13}$CO}

\def\etal {et al.}
\def\kms {\hbox{${\rm km\,s}^{-1}$}}

\def\edcomment#1{\iffalse\marginpar{\raggedright\sl#1\/}\else\relax\fi}
\reversemarginpar

\begin{document}

\title{An Inner Molecular Bar or Disk in NGC~5195}
 \author{S. Aalto}
 \author{G. Rydbeck}
\affil{Onsala Rymdobservatorium, Chalmers Tekniska H\"ogskola,
S-439 92 Onsala, Sweden}

\begin{abstract}
High resolution OVRO CO 1-0 observations of the inner 
kpc of the M51-companion NGC~5195 reveal the
presence of a kpc-sized bar, or possibly an inclined disk with
a two armed spiral, at the center of the optical bar. 
The molecular mass of the feature is $2.7 \times 10^8$ M$_{\odot}$, half
of which is within a radius of 250 pc. The resulting gas surface density of
$10^3$ M$_{\odot}$ pc$^{-2}$ is typical for starbursts.
However, the lack of evidence for current star formation suggests that either
some mechanism is preventing
stars from forming, or the standard CO to H$_2$ conversion factor
substantially overestimates the available amounts of molecular gas. 
\end{abstract}

\section{Introduction}

NGC~5195, the intriguing SBa/0 companion of M51, shows exceptionally bright 
15 $\mu$m emission from a central point source suggesting intense nuclear
activity (Boulade \etal\ 1996). Warm IR colours and a respectable FIR
luminosity points towards a
starburst scenario, while the lack of bright H$\alpha$ emission seems to 
contradict this notion. Boulade \etal\ suggest that the activity is powered by
an evolved starburst, but the possibility of LINER activity or ongoing star
formation is also considered.
High resolution CO observations help us address both the nature of the
central activity as well as the mechanisms responsible for feeding the nucleus
with molecular gas. Secondary features, such as bars within bars, have been proposed
as a possible mechanism of gas fueling nuclear activity (e.g. Shlosman \etal\ 1990). 
The proximity of NGC~5195
allows detailed studies of gas transport processes that are likely to feed the
activity also within more distant, luminous systems.
Below we present a preliminary analysis of our high resolution OVRO CO 1-0 observations.

\section{Molecular gas distribution and kinematics}

In Figure 1 we see the CO distribution in the eastern and central part of
NGC~5195 overlayed on a DSS optical image. Even though we have not mapped the 
whole of NGC~5195 it is clear that the gas is very concentrated towards the
center. The eastern CO emission is well correlated with the
foreground dust absorption in the M51 northern arm which covers part of the galaxy. 
The inner 500 pc of NGC~5195 is in rough solid body rotation with a position angle of 
$\approx 95^{\circ}$. The projected maximum velocity is 80 \kms, and
the dynamical mass inside a radius of 250 pc is $7 \times 10^8$ M$_{\odot}$ 
(for $i = 45^{\circ}$).
Figure 2 shows higher resolution images of the central region of NGC~5195. 
The CO emission is double
peaked and appears to be part of a ring-like structure. The kinematic
center is however not at the center of this ring, but rather on the brighter
of the two peaks, which also coincides with the NIR peak emission. The peak brightness temperature
is 5 K (for a linear resolution of $\approx 100$ pc). This suggests that the brightness
temperature of each cloud is rather moderate, or that there is an ensemble of warm
clouds of low filling factor. The morphology of the CO emission suggest that the gas clouds
could be travelling along the x2 orbits of the bar. The possibility of an inclined disk with
spiral arms cannot, however, be excluded at this stage.  Disturbances in the velocity field
indicate the presence of gas streaming in the inner kpc.


\begin{figure}
\plotone{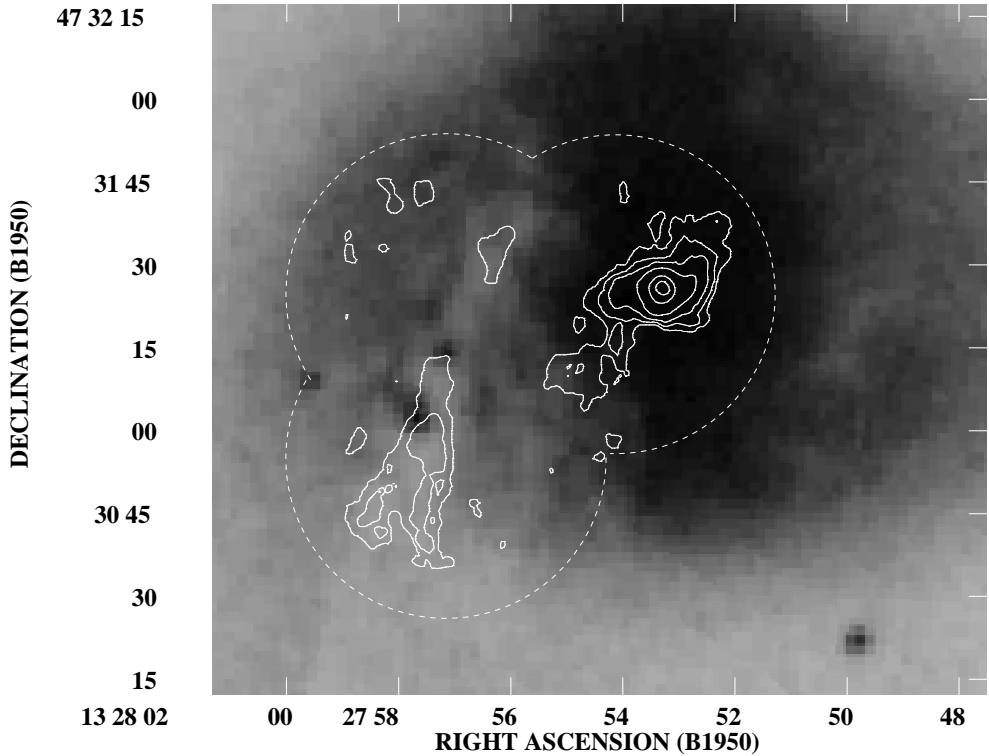}
\caption{Contours show the integrated CO intensity (resolution 4$''$) overlayed
on a grayscale DSS image of NGC~5195. The contour levels are 
(1,3,6,12,24,30)$\times$ 2 Jy beam$^{-1}$ \kms. The dashed line marks the area
(three pointings) mapped by OVRO.}
\end{figure}


\subsection{Molecular gas mass}

The total recovered flux in NGC~5195 is 320 Jy \kms, corresponding to $2.7 \times 10^8$
M$_{\odot}$ for $D$=9.6 Mpc (for a conversion factor of $X$ = N(H$_2$)/I(\twco)
= $2.3 \times 10^{20}$ $\cmmt$ (K \kms)$^{-1}$)). Half of this gas mass is found within
a radius of 250 pc. This results in an average gas surface density close to 
$10^3$ M$_{\odot}$ pc$^{-2}$ in the inner 500 pc, and 20\% of the dynamical mass is molecular.  
For the M51 arm we find 81 Jy \kms\
corresponding to a mass of $6.8 \times 10^7$ M$_{\odot}$. 
We recover 60\% $\pm$ 20\% of the OSO 20m single dish flux in the same area.

\begin{figure}
\plottwo{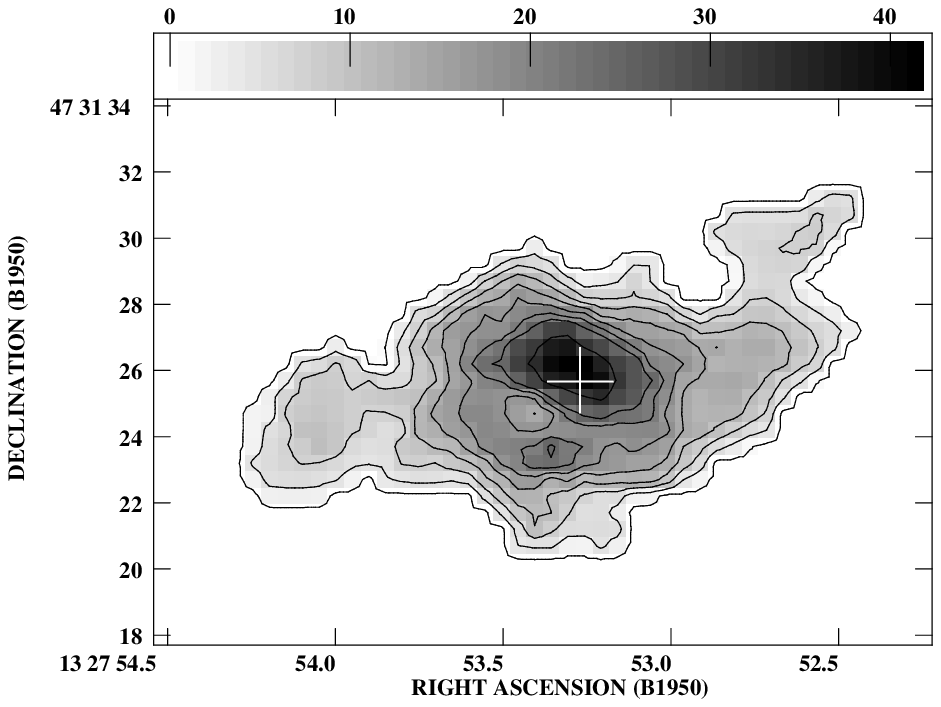}{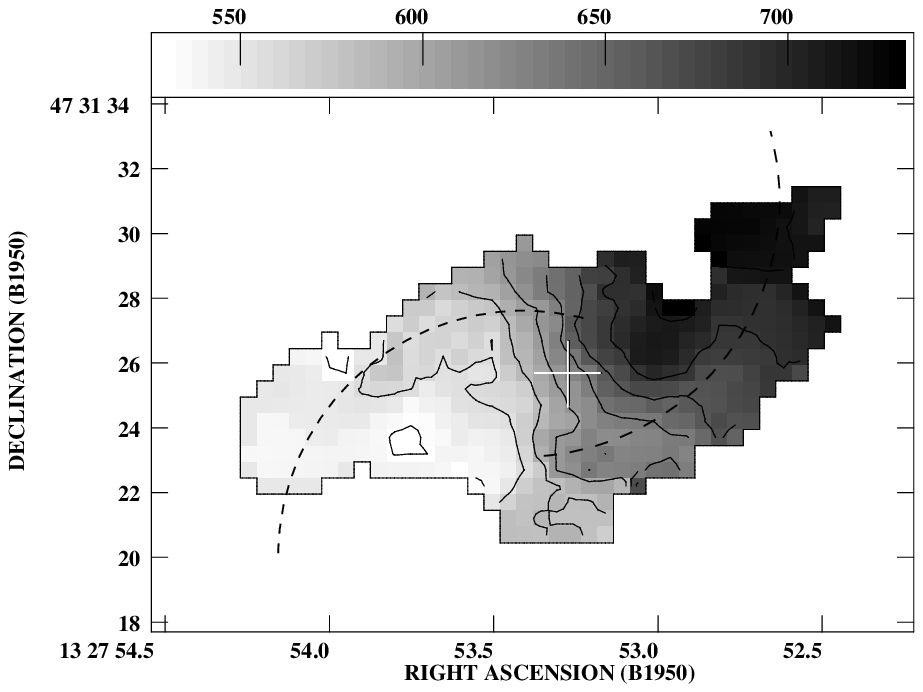}
\caption{The left panel shows the high resolution (2$''$) CO 1-0 integrated
intensity image. The grayscale ranges from 0 to the peak flux
(41.4 Jy beam$^{-1}$ \kms) and the contours go from
(1,5...29,40,...,90)$\times$ 0.83 Jy beam$^{-1}$ \kms.
 The right panel shows the velocity field where the contours range from 550 \kms\
 to 740 \kms\ with 27 \kms\ increment. The grayscale range from 530 to 730 \kms. 
 The white cross marks the position of
 the peak 2.2 $\mu$m emission (Smith \etal\ 1990). The dotted lines in the right figure
 mark the suggested position of the inner spiral arms.}
\end{figure}

\section{Discussion}

\subsection{The starburst}

The highly concentrated CO emission of NGC~5195 is fairly typical of interacting systems,
where the central molecular gas fuels starburst activity and/or an AGN.
NGC~5195 is, however, classified as an {\it old} starburst where star formation ceased
some $10^7$ years ago, and
the absence of bright H$\alpha$ emission supports this notion. Diffuse, shell-like H$\alpha$
features may indicate a nuclear outflow similar to that of M~82 (Greenawalt 1998).
It is interesting, therefore, to ponder why the previous starburst left a considerable amount
of gas behind in the center. 
What mechanism made the star formation processes grind to a halt before the gas was consumed?
The inner region of the
galaxy is in an apparent solid body rotation, which should be favourable for star formation.
The surface density of gas left behind is typical of that found in starburst
galaxies, again offering no direct solutions to the cessation of star formation. It is of course
possible
that gas has been transported to the center after the previous burst of star formation,
but that still leaves us with the question of what is preventing the gas from forming stars.

A possible clue may lie in the excitation of the molecular gas itself. From our single dish
study of NGC~5195 we find that the emission from \thco\ 1-0 is unusually faint with respect
to the \twco\ emission. The \twco/\thco\ 1--0 intensity ratio is $\gapprox 20$, while the more
typical value for galaxies is 10-15 (e.g. Aalto \etal\ 1995). The elevated ratio is caused by
an overall reduction in the optical depth of the \twco\ line which would likely be caused
either by a) high temperatures in the gas
or b) presence of diffuse unbound molecular gas. To distinguish between the two scenarios
it is necessary to observe higher transitions of both \twco\ and \thco. Both scenarios may
lead to an overestimate of the
amount of molecular gas present by factors of 5-10. In this case, the starburst may simply have
ceased because of the gas surface density being below threshold values. 
Smith (1982) find a high dust temperature ($T_{\rm d} \approx 65$ K) for NGC~5195 which is
comparable to the warmest of the luminous starburst galaxies. If indeed the activity in NGC~5195
is dominated by an evolved starburst, then the question is how it is capable of heating the
dust to such high temperatures. 

\subsection{The inner bar or disk}

The inner elongated structure is oriented roughly at right angles to the larger scale
optical bar. This feature has not been seen before in NIR or optical, even if a $J$-band
elongation at lower resolution (Smith \etal\ 1990) may well be associated with it.
The discontinuities in the velocity field (see Figure 2) suggest the presence of
gas streaming in the inner kpc of NGC~5195. Such regions 
are often associated with dust lanes and downstream 
H$\alpha$ emission, but only very faint and diffuse H$\alpha$ emission can be found in
NGC~5195 and the shock may be too weak to produce sharp dust lanes.
Kenney \etal\ 1992 discuss ``twin peaks'' features perpendicular to optical bars as
resulting from orbit crowding near the ILR (Inner Lindblad Resonance). We speculate that
the molecular feature in NGC~5195 shows that the gas has passed beyond the ILR to the nucleus
of the galaxy. Further study, of both the overall kinematics
of NGC~5195 as well as higher resolution observations of CO and NIR, is necessary to establish
the true nature of the central molecular concentration.



\end{document}